\documentclass[12pt]{iopart}

\usepackage{iopams}  

\usepackage{graphicx}
\usepackage{inputenc}[utf8]
\usepackage{braket}
\usepackage{array}
\usepackage{siunitx}
\usepackage[numbers, sort&compress]{natbib}
\usepackage[colorlinks=true,linkcolor=blue,citecolor=blue,urlcolor=blue,pdfborder={0 0 0}]{hyperref}

\newcommand{\abs}[1]{\left| {#1} \right|}
\renewcommand{\d}{\mathrm{d}} 
\renewcommand{\t}{\text} 

\renewcommand{\Tr}{\mathrm{Tr}} 
\renewcommand{\e}{e}

\newcommand{\rs}{\scriptscriptstyle\mathrm}

\renewcommand{\vec}[1]{\mathbf{#1}}

\newcommand{\ketbra}[1]{\ket{#1}\hspace{-1mm}\bra{#1}}

\begin{document}

\title{Quantum Optics with Rydberg Superatoms}

\author{Jan Kumlin$^1$, Christoph Braun$^{2,3,4}$, Christoph Tresp$^5$\footnote{Current address: TOPTICA Photonics AG}, Nina Stiesdal$^6$ , Sebastian Hofferberth$^6$, Asaf Paris-Mandoki$^7$\footnote{Email: asaf@fisica.unam.mx}}

\address{${}^1$ Center for Complex Quantum Systems, Department of Physics and Astronomy, Aarhus University, Ny Munkegade 120, DK-8000 Aarhus C, Denmark, ${}^2$ \,Fakult\"at f\"ur Physik, Ludwig-Maximilians-Universit\"at M\"unchen, Schellingstra{\ss}e 4, 80799 M\"unchen, Germany, ${}^3$\,Max-Planck-Institut f\"ur Quantenoptik, Hans-Kopfermann-Stra{\ss}e 1, 85748 Garching, Germany, ${}^4$\,Munich Center for Quantum Science and Technology (MCQST), Schellingstr. 4, 80799 München, Germany, ${}^5$ 5. Physikalisches Institut and Center for Integrated Quantum Science and Technology,
Universität Stuttgart, Pfaffenwaldring 57, 70569 Stuttgart, Germany, ${}^6$ Institute of Applied Physics, University of Bonn, Wegelerstraße 8, 53115 Bonn, Germany, ${}^7$ Instituto de Física, Universidad Nacional Autónoma de México, 04510 Mexico City, Mexico}
\vspace{10pt}
\begin{indented}
\item[]February 2023
\end{indented}

\begin{abstract}
Quantum optics based on highly excited atoms, also known as Rydberg atoms, has cemented itself as a powerful platform for the manipulation of light at the few-photon level. The Rydberg blockade, resulting from the strong interaction between individual Rydberg atoms, can turn a large ensemble of atoms into a system which collectively resembles a single two-level emitter, a so-called Rydberg superatom. The coupling of this artificial emitter to a driving photonic mode is collectively enhanced by Rydberg interactions, enabling strong coherent coupling at the few-photon level in free-space. The exquisite level of control achievable through this has already demonstrated its utility in applications of quantum computing and information processing. Here, we review the derivation of the collective coupling between a Rydberg superatom and a single light mode and discuss the similarity of this free-space setup to waveguide quantum electrodynamics systems of quantum emitters coupled to photonic waveguides. We also briefly review applications of Rydberg superatoms to quantum optics such as single-photon generation and single-photon subtraction.
\end{abstract}

%
%
%
%
%

\section{Introduction}

A non-linear optical medium can be understood as having an index of refraction which depends on the intensity of incoming light~\cite{boydNonlinearOptics2008}. When the nonlinearlity is strong enough to be significant at the few-photon level such media can be used to selectively control light based on the number of photons within it enabling the preparation of non-classical states of light~\cite{Vuletic2012,Hofferberth2016d} and the implementation of optical quantum gates~\cite{Duerr2019}. There are several strategies actively being pursued~\cite{Chang2014} to achieve the few-photon level of optical nonlinearity. These strategies ranges from cavity quantum electrodynamics (QED)~\cite{Birnbaum2005} with single quantum emitters, where the coupling is enhanced by the repeated interaction of single photons with the emitter, to systems where many quantum emitters collectively modify the single photon transmission~\cite{solanoOpticalNanofibers2017,sheremetWaveguideQuantumElectrodynamics2023}. The focus of this paper is to discuss a type nonlinear optical media based on atoms in highly excited atomic states, also known as Rydberg atoms where, as it will be shown in this review, the collective nature of an excitation can result in a directed emission which can be directly understood as a waveguide QED system. 

Atoms in Rydberg states have several properties~\cite{Gallagher1994,Low2012a} which are significantly different from ground-state atoms and are relevant for quantum optics. As the principal quantum number $n$ increases, the expected value of the atomic radius grows as $\left\langle r \right\rangle \sim n^2$ such that for sufficiently large $n$ even several atoms can fit inside the wavefunction of the Rydberg electron~\cite{Gaj2014}. A very consequential scaling is that of the transition dipole moments. For instance, the scaling of the transition dipole moment from the ground state to a Rydberg state is $\left\langle nl |er|g \right\rangle \sim n^{-3/2}$ results in a significantly increased radiative lifetime $\tau$ scaling as $\tau\sim n^3$.  On the other hand, the transition dipole moment between adjacent Rydberg states scales as  $\left\langle nl |er|nl' \right\rangle \sim n^2$. As a result, and taking into account that the scaling of the level spacing between adjacent states is $\Delta E_\mathrm{adj} \sim n^{-3}$, it is straightforward to find through a second-order perturbation calculation that atomic polarizability scales as $\alpha \sim n^7$. This outcome is expected because a wavefunction that is more spread out has less binding energy than a wavefunction that is tightly localized around the nucleus. Hence, the impact of an external electric field is more pronounced for higher values of $n$. Finally, even though atoms in a Rydberg state $\ket{nl}$ have zero dipole moment, two atoms may interact via a second-order effect of dipole-dipole interaction~\cite{Hofferberth2017, Weatherill2017} resulting in a van der Waals interaction of the form $V(R) = C_6/R^6$ for which $C_6 \sim n^{11}$.

Among the many fascinating aspects of highly excited Rydberg states~\cite{Gallagher1994,Saffman2010,Shaffer2019}, one feature that has emerged as a key concept for quantum information and simulation is the Rydberg blockade. When a laser drives an ensemble of atoms towards a Rydberg state, the strong interaction between them leads to the Rydberg blockade phenomenon. This means that within a certain distance, known as the Rydberg blockade length, from a first Rydberg atom, the excitation of a second or subsequent atoms is prevented~\cite{Lukin2001c}. The Rydberg blockade has been observed in ultracold atomic systems in the frozen Rydberg-gas regime~\cite{Gallagher1998,Pillet1998}, both in bulk ensembles~\cite{Weidemueller2004,Gould2004,Raithel2005,Pillet2006,Pfau2007b,vanLinden2008,Raithel2008,Pfau2008,Weidemueller2008,Entin2010,Raithel2011} and in small systems supporting only a single excitation~\cite{Kuzmich2012c,Walker2014,Ott2015,Gross2015c}. Also, experiments to probe Rydberg interaction effects in room-temperature thermal vapors have been carried out~\cite{Pfau2013b,Adams2013b,Loew2015}. 

The Rydberg blockade mechanism has been exploited to realize atomic two-qubit gates~\cite{Saffman2009,Grangier2009,Grangier2010,Saffman2010b}, achieving fidelities $\geq 0.991(4)$ ~\cite{madjarovHighfidelityEntanglementDetection2020} using alkaline-earth atoms, as basic components of large-scale neutral atom quantum registers~\cite{Saffman2015,Saffman2015b}. In combination with single-layer optical lattices~\cite{Greiner2009,Kuhr2010,Bloch2012} or configurable single-atom tweezer arrays~\cite{Browaeys2016b,Browaeys2016d,Lukin2016c,Regal2015,Ahn2016,Kaufman2018,Endres2019,Thompson2019,wangPreparationHundredsMicroscopic2020}, Rydberg blockade enables the quantum simulation of large interacting spin systems~\cite{Gross2015,Lukin2017d,Gross2017b,Browaeys2018,Ahn2018,Lukin2019b,Lukin2019d,Browaeys2019,bluvsteinQuantumProcessorBased2022}, innovative nonlinear optical media with greatly suppressed losses~\cite{moreno-cardonerQuantumNonlinearOptics2021, zhangPhotonphotonInteractionsRydbergatom2022}, and novel approaches for implementing quantum computers based on the Rydberg blockade have been proposed~\cite{liuNonadiabaticNoncyclicGeometric2020, cohenQuantumComputingCircular2021,congHardwareEfficientFaultTolerantQuantum2022,wuUnselectiveGroundstateBlockade2022}.

When applied to atomic ensembles instead of small numbers of individual atoms, the Rydberg blockade leads to the creation of Rydberg superatoms, where a large number of atoms share a single Rydberg excitation~\cite{Lukin2001c} and undergo collectively enhanced Rabi oscillations when driven by external fields~\cite{Kuzmich2012c,Ott2015,Gross2015c,Browaeys2016b}. From a quantum optical viewpoint, a Rydberg superatom acts as a single two-level quantum emitter with a collectively enhanced coupling to the forward direction of the excitation field mode~\cite{Kuzmich2012b,Hofferberth2017c}, which has been exploited for highly efficient single-photon generation~\cite{Kuzmich2012b}, entanglement generation between light and atomic excitations~\cite{Kuzmich2013} and as a collectively encoded qubit~\cite{spongCollectivelyEncodedRydberg2021,xuFastPreparationDetection2021,meiTrappedAlkaliMetalRydberg2022}. Furthermore, the single-photon generation scheme has been successfully implemented in a room-temperature atomic vapor cell~\cite{Pfau2018b}. To account for the internal structure of the collective excitation, an additional set of dark states which do not couple to the driving light but can couple to the single bright state through dephasing processes must be included~\cite{Buechler2011}. For large number of atoms $N$ composing a single superatom, the collective enhancement of the coupling can become sufficiently strong so that the interaction of individual photons with this artificial emitter can be studied in free-space without any resonator or waveguide structures to enhance the atom-light interaction~\cite{Hofferberth2017c,Hofferberth2018}.

Employing Rydberg superatoms to manipulate optical photons is intricately related to Rydberg polaritons, where the Rydberg interactions in atomic gases are mapped onto photons by means of electromagnetically induced transparency (EIT)~\cite{Pohl2016d,Hofferberth2016d} to realize few-photon optical nonlinearities~\cite{Kurizki2005,Adams2010,Lukin2011,Kurizki2011,Pohl2013,Vuletic2012,Weidemueller2013}. Rydberg polaritons have been exploited to realize single-photon all-optical switches~\cite{Duerr2014} and transistors~\cite{Hofferberth2014,Rempe2014b,Hofferberth2016},  interaction-induced phase shifts~\cite{Grangier2012,Duerr2016,Vuletic2017}, conditioned single-photon deflection~\cite{Adams2017} and most spectacularly to implement a full two-photon quantum gate~\cite{Duerr2019}. Additionally, Rydberg EIT provides access to novel phenomena such as attractive interaction between single photons~\cite{Vuletic2013b}, crystallization of photons~\cite{Fleischhauer2013}, or photonic scattering resonances~\cite{Buechler2014}.

More recently, efforts directed at exploring Rydberg superatom systems contained within optical cavities~\cite{suarezSuperradianceDecoherenceCaused2022,jiaStronglyInteractingPolaritonic2018a, Boddeda2016} have increased. One particular direction being explored focuses on creating a qubit that can be addressed individually. In this approach a single superatom is used to ensure that, at most, one excitation can be produced in the system and preventing the accidental excitation of a secondary qubit~\cite{ghoshCreatingHeraldedHyperentangled2021, vaneeclooIntracavityRydbergSuperatom2022, yangDeterministicMeasurementRydberg2022, yangSequentialGenerationMultiphoton2022, stolzQuantumLogicGateTwo2022}. The cavity enhances the superatom-photon coupling which, together with the collective enhancement due to the blockade, allows coherent control and faithful measurement of the superatom qubit~\cite{vaneeclooIntracavityRydbergSuperatom2022, yangDeterministicMeasurementRydberg2022}.  The superatom-cavity approach has made it feasible to implement new types of quantum gates for optical photons~\cite{stolzQuantumLogicGateTwo2022}, build sources of non-classical states of light~\cite{yangSequentialGenerationMultiphoton2022,magroDeterministicFreePropagatingPhotonic2022}, engineer novel quantum matter~\cite{Clark2019}, and even realize topological states of light~\cite{clarkObservationLaughlinStates2020} that were previously only observed in solid-state electron systems.

In this tutorial, we discuss the basic aspects of a single Rydberg superatom, formed by a large number of three-level atoms, $N$, all within a fully blockaded volume driven by a weak probe and strong control laser fields. We review the theoretical steps that show how this system of $N$ three-level atoms can be reduced to a single effective two-level system. Then we proceed to discuss the interaction of this single superatom with a quantized light mode. Finally, we briefly review two quantum optical applications of a single superatom, namely single-photon generation and single-photon subtraction.

\begin{figure}
\centering
	\includegraphics{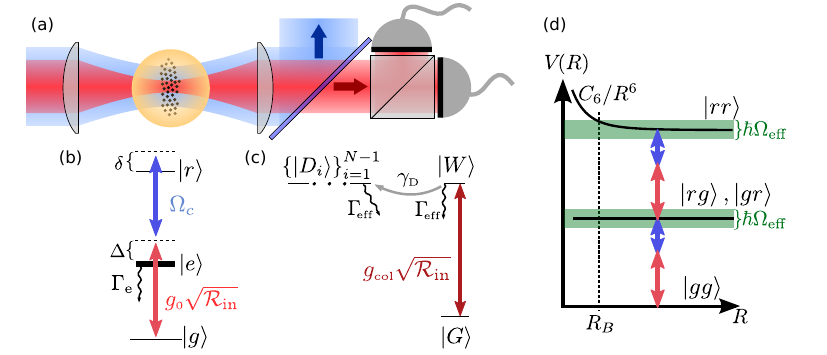}
		\caption{(a) Experimental scheme for preparation and addressing of a Rydberg superatom. The tightly confined cloud resides within an optical dipole trap confining it to a volume smaller than $V_B$, the control field is diverted from the photon counting modules detecting the transmitted photon field. (b) Level scheme of a single atom. The atom is addressed with two laser beams, coupling the $\ket{g}$ to $\ket{r}$ via the intermediate state $\ket{e}$. (c) Collective levels of the Rydberg superatom. The collective ground state is coherently coupled to the collective excited state, which is incoherently coupled to the manifold of dark states. (d) Interaction energy for a pair of atoms as a function of interatomic distance $R$. At distances smaller than $R_B$ the interaction detunes the energy level $\ket{rr}$ from the excitation lasers so that a second excitation cannot occur.   }
        \label{fig:system}
\end{figure}

\section{Theoretical description of the superatom}
\label{sec:superatom}

In the following we consider an ensemble of $N$ three-level atoms, which is addressed by two laser fields. For convenience the three atomic states involved will be abbreviated by $\ket{g}$ for the ground state, $\ket{e}$ for the intermediate state, and $\ket{r}$ for the Rydberg state. The Rabi frequencies will for now be assumed to be constant in time and labeled $\Omega_p = g_0 \sqrt{R_{in}}$ and $\Omega_c$ for the transitions $\ket{g} \leftrightarrow \ket{e}$ and  $\ket{e} \leftrightarrow \ket{r}$ respectively, Figure~\ref{fig:system}(b). The Rabi frequency on the lower transition is determined by a coupling constant $g_0$, given by the spatial geometry of the probe field and the incoming photon flux $\mathcal{R_\text{in}}$; we consider the upper transition to be driven by classical light. We also consider a decay from $\ket{e}$ to $\ket{g}$ with rate $\Gamma_\mathrm{e}$.

In order to get an understanding of the interesting physics that are enabled by collective effects of interacting Rydberg atoms, it is convenient to summarize the essential features of the Rydberg blockade mechanism.

Atoms in the Rydberg state $\ket{r}$ interact strongly with each other, the strong interaction arises from the long-wavelength dipole-dipole interactions between their constituents~\cite{Hofferberth2017,Weatherill2017, Lukin2001c, Gallagher1994}. When the atoms are far enough apart, such that their electronic wavefunctions do not overlap~\cite{LeRoy1973} the interaction energy can be written as a multipole expansion. 
This interaction-Hamiltonian modifies the energy level of a pair of atoms in the state $\ket{rr}$ and depends on the distance $R$ between the atomic wavefunctions, thus the interaction potentials may be calculated as a function of $R$ employing the Born-Oppenheimer approximation. Depending on the detailed level structure of the atom pair their interaction can be approximated for sufficiently large separations $R$ to have a $1/R^3$ or a $1/R^6$ dependence~\cite{Hofferberth2017}. Nonetheless, the exact shape of the potential is of little relevance to understand the blockade effect we are interested in. For concreteness we will consider the most common case in experiments, where the energy of a pair of atoms in the Rydberg state has the form $V(R)=C_6/R^6$ as in Figure~\ref{fig:system}(d). Here, $C_6$ reflects the strength of the interaction. For two atoms at a distance $R$ the two laser fields will excite both atoms only, if they are further separated than the blockade radius $R_B=(\hbar \Omega_\t{eff}/C_6)^{-1/6}$, $\hbar$ is the reduced Planck constant and $\Omega_\t{eff}$ is the effective excitation bandwidth between the ground state and the Rydberg state of the system. As the $C_6\propto n^{11}$ coefficient scales dramatically with the principal quantum number $n$, the blockade radius enables mesoscopic distances to be bridged. E.g. for a ${}^{87}$Rb atom in the state $\ket{121S_{1/2}, m_J=1/2}$ and the effective Rabi frequency reported in~\cite{Hofferberth2016e}, the blockade radius is $R_B\approx\SI{17}{\micro\meter}$. 

Having a mechanism available that effectively prevents the simultaneous excitation of multiple Rydberg atoms within the blockade volume $V_B = 4\pi R_B^3/3$, allows for the creation of highly entangled collective quantum states inside this volume. The specific setup we want to consider is a cold, but not quantum-degenerate, cloud of atoms that is well smaller than the blockade volume, Figure~\ref{fig:system}(a). Thus only a single Rydberg atom can be present in the entire ensemble. If the cloud is collectively addressed, the single allowed excitation is distributed among all constituents. 

The objective of this section is to detail the interaction of a Rydberg superatom with a free-space Gaussian  mode and to understand how the strong coupling to the Rydberg level with a few-photon field comes about. For an atom to strongly couple to a photon in free space the figure of merit is the line-width or scattering rate of a given transition. For the alkali atoms, the transitions from ground to the first-excited states (also known as $D$-lines) are the strongest~\cite{bransdenPhysicsAtomsMolecules2003}, yet they do not connect to Rydberg states. Therefore the excitation to the Rydberg state is done via a two-photon-transition with the aid of an intermediate state. We begin by showing in~\ref{sec:three-levels-turned-adiabatic-elimination} that by choosing appropriate parameters, the intermediate state is never significantly populated, therefore irrelevant for the dynamics of the individual atoms, and can be removed from the description through so-called  ``adiabatic elimination''. This is followed by section \ref{sec:collective_emission}  where we begin describing the interaction of an atomic cloud with a single plane-wave photon and finally turn our attention to consider a Gaussian mode.

\subsection{Adiabatic Elimination of the single-atom intermediate state}
\label{sec:three-levels-turned-adiabatic-elimination}

Excitation to a Rydberg state is often done via an intermediate state as in Figure~\ref{fig:system}(b), since it can enable larger coupling compared to a direct transition e.g. in the UV-range of the spectrum. For example, in $^{87}$Rb atoms, the ground state $\ket{5S^{1/2}, F=2, m_F=2}$ can be coupled via the intermediate state $\ket{5P^{3/2}, F=3, m_F=3}$ to any $nS$ or $nD$ Rydberg state~\cite{Low2012a}. This offers the opportunity to utilize the large dipole matrix element of the $D_2$-line in alkali atoms and compensate for the smaller transition probability to the Rydberg state with an increased Rabi-frequency, via high power lasers, on the transition from the intermediate to the Rydberg state. As will be shown below, The higher Rabi frequency $\Omega_c$ of the upper transition does not only increase the coupling but also the decay via the intermediate state. A direct transition via highly energetic UV-photons to a $nP$ Rydberg state  offers less flexibility as the effective Rabi frequency cannot be adjusted via the intermediate state detuning $\Delta$. Additionally, building optical arrangements to work with UV is more challenging than for light that falls within the visible spectrum.

When there is no cavity or waveguide modifying the mode structure of the electromagnetic field all fields correspond to modes in free space. 
The coherent evolution of the three levels within the rotating wave approximation in the rotating frame can then be described by the Hamiltonian 
\begin{equation}
  H_{3L}= \hbar\left(
    \begin{array}{ccc}
    0 & \Omega_p/2 & 0 \\
    \Omega_p^*/2 & -\Delta & \Omega_c/2 \\
    0& \Omega_c^*/2 &  - \delta 
  \end{array}\right),
\end{equation}
where $\Delta=\omega_p-(E_e-E_g)/\hbar$ is the detuning for the transition from $\ket{g}$ to $\ket{e}$,  $\delta=\omega_p+\omega_c-(E_e-E_g)/\hbar-(E_r-E_e)\hbar$ is the two-photon detuning and $\Omega_p$, $\Omega_c$ are the Rabi frequencies for the lower and upper transitions respectively as in Figure~\ref{fig:system}(b).
The time-evolution governed by this Hamiltonian for any state
\begin{equation}
  \Psi(t) = c_g(t) \ket{g} + c_e(t) \ket{e} + c_r(t) \ket{r}
\end{equation}
where $\ket{g}$, $\ket{e}$ and $\ket{r}$ are time-independent state vectors, is determined by the Schrödinger equation which yields three coupled differential equations for the coefficients
\begin{eqnarray}
  i \partial_t c_g(t) &= \frac{\Omega_p}{2} c_e(t) \\
  i \partial_t c_e(t) &= -\Delta c_e(t) + \frac{1}{2}\left( \Omega_c c_r(t) + \Omega_p^* c_g(t) \right) \\
  i \partial_t c_r(t) &= -\delta c_r(t) + \frac{\Omega_c^*}{2} c_e(t).
\end{eqnarray}

If the laser-fields are sufficiently far detuned from the intermediate state, i.e. $\Delta \gg \Omega_p, \Omega_c$, and the two photon detuning $\delta$ is small, the intermediate state almost immediately follows all dynamics, such that it always equilibrates compared to the timescale of the other, slower dynamics. The two resulting timescales allow for separation of the fast and slow dynamics, which is commonly referred to as \emph{adiabatic elimination}. In the aforementioned limit of large detuning the dynamics of the intermediate state correspond to high frequency oscillations with small amplitude, compared to slower oscillations occurring for $\ket{g}$ and $\ket{r}$. For sufficiently large detuning the dynamics of the intermediate state can thus be eliminated, as it does not influence the dynamics of interest, i.e. the evolution of $\ket{g}$ and $\ket{r}$. 

This can be seen by considering the dynamics related to two sub-spaces: one defined by the projector $P=\ketbra{g}+\ketbra{r}$ involving the states that participate in the dynamics of interest, occurring on a slow timescale compared to the fast dynamics of the Hamiltonian $H$, and its complementary subspace given by the projector $Q=\ketbra{e}$ such that the sum of the subspaces describes the entire system and therefore $P+Q=1$. The eigenvalues of $QHQ$ are widely separated from those of $PHP$, and the coupling between the two subspaces is small compared to the eigenvalues of $QHQ$~\cite{Egusquiza2016}. Starting from the Schrödinger equation one finds
\begin{eqnarray}
  i \hbar \partial_t (P+Q) \Psi(t) &= (P+Q)H(P+Q)\Psi(t)\\
  i \hbar \partial_t (P\Psi(t)) &= PHP (P\Psi(t)) + PHQ (Q\Psi(t)) \\
  i \hbar \partial_t (Q\Psi(t)) &= QHP (P\Psi(t)) + QHQ (Q\Psi(t)),
\end{eqnarray}
where the last two equations are obtained by multiplying with $P$ and $Q$ respectively and using $P^2=P$ and $Q^2=Q$. If the system now starts from $P \Psi(t)$, since the interaction between the subspaces is weak, the wave function of $Q \Psi(t)$ will not change significantly. Therefore setting $\partial_t Q\Psi(t) = 0$ is justified, which results in~\cite{Egusquiza2016}
\begin{equation}
  i \partial_t (P \Psi(t)) = PHP (P\Psi(t)) - PHQ (QHQ)^{-1} QHP (P\Psi(t)).
\end{equation}
Applying this to $H_{3L}$ means setting $\partial_t c_e(t)=0$, solving for $c_e(t)$ and replacing it in the remaining equations to obtain~\cite{brionAdiabaticEliminationLambda2007}
\begin{eqnarray}
  i \partial_t c_g(t) &= \frac{\abs{\Omega_p}^2}{4\Delta}c_g + \frac{\Omega_p \Omega_c}{4\Delta} c_r(t)\\
  i \partial_t c_r(t) &= \frac{\Omega_p^* \Omega_c^*}{4\Delta} c_g(t) + \left(\frac{\abs{\Omega_c}^2}{4\Delta}- \delta\right) c_r(t). 
\end{eqnarray}
The dynamics of this system correspond to an effective two-level system where the levels are coupled by a driving field with Rabi frequency $\Omega_\text{eff} = \Omega_p \Omega_c/(4\Delta)$ and effective detuning $\delta_\text{eff}=\delta+\left(\abs{\Omega_p}^2-\abs{\Omega_c}^2\right)/(4\Delta)$. The intermediate state's population, when solving the full equations, is small at all times and thus does not contribute significantly to the dynamics.

Considering the problem from an experimental point of view, each atom, though well controlled, cannot be perfectly isolated and thus decays due to its coupling to the environment. The previous treatment did not include any decay channels of the individual excited states. 
Even though the population in $\ket{e}$ is negligible, it still contributes to the decay of $\ket{r}$. The state $\ket{r}$ is a long-lived Rydberg state whereas the intermediate state $\ket{e}$ exhibits a non-negligible decay rate $\Gamma_\text{e}$. Thus the contribution of $\ket{e}$ to $\ket{r}$ due to the large  Rabi frequency of the coupling field $\Omega_c$ between $\ket{e}$ and $\ket{r}$, leads to a decay channel for the Rydberg state. The admixture can be calculated by either diagonalizing the Hamiltonian $H_{3L}$ and approximating the eigenstates for $ \Omega_p, \Omega_c, \delta \ll \Delta$, or calculated directly with first order perturbation theory and considering $\delta \ll \Delta$, which gives 
\begin{equation}
   \ket{r'} \approx \ket{r} + \frac{\Omega_c}{2\Delta}\ket{e}.
\end{equation}
This results in an effective decay rate $\Gamma_\text{eff}$ of the population in $\ket{r}$ due to the admixture of $\ket{e}$ and it's decay to $\ket{g}$ with rate $\Gamma_\mathrm{e}$ given by
\begin{equation}
  \Gamma_\text{eff} = \frac{\Omega_c^2}{(2\Delta)^2} \Gamma_\mathrm{e}.
\end{equation}

The linewidth of the coupled system can in principle be as narrow as the linewidth of the uppermost excited state, for $\abs{\Omega_c/\Delta} \ll 1$. When utilizing a strong effective Rabi frequency, the effective decay rate often provides the dominant term compared to the bare Rydberg lifetime.

\subsection{Collective Excitation and Emission}
\label{sec:collective_emission}

As discussed earlier, once an atomic ensemble is smaller than the blockaded volume, it can only host a single Rydberg excitation. In this section we want to discuss the cooperative effects resulting from the Rydberg blockade. In the following the case of a single photon interfacing a large ensemble of emitters will be investigated. This setting is also discussed in great detail in~\cite{Scully2008, Chang2008, Scully2010, Scully2007, Wodkiewicz2006, Scully2010b}. A more extensive treatment can be found in~\cite{Lehmberg1970, Cirac2008} also including multiple photons without blockade.

The reduction to consider only a single photon addressing an ensemble of atoms is done in order to simplify the setting and illustrate the main consequences. In the presence of an interaction mechanism (e.g. Rydberg blockade), that prevents multiple excitations within the ensemble, also multiple photons are only scattered one at a time. Therefore first considering only a single photon interacting with the cloud will illustrate the main features of the system. Note that this description cannot include stimulated emission as this is a multi-photon effect.

The operator describing the interaction of an atomic cloud with an incident plane wave electric field is given by the sum of all single atom operators~\cite{Wodkiewicz2006}
{
\setlength{\mathindent}{0cm}
\begin{equation}
  V(t) = \sum_j \hbar g_0 \left( \sigma^\dagger_j a_\mathbf{k_0}  \e^{i \mathbf{k_0}\cdot\mathbf{r}_j} \e^{-i(\nu_0 -\omega_0)t} +  \sigma_j a^\dagger_\mathbf{k_0}  \e^{-i \mathbf{k_0}\cdot\mathbf{r}_j} \e^{i(\nu_0-\omega_0)t}\right) + \frac{1}{2}\sum_{j\neq i} V_{i,j} \sigma^\dagger_j\sigma_j\sigma^\dagger_i\sigma_i
\end{equation}
}
where $\sigma_j= \ket{g}_j\bra{r}_j$ is the single atom operator relaxing atom $j$ from the excited to the ground state  and $a_\mathbf{k_0}$ the annihilation operator of a photon in mode $k_0$ with frequency $\nu_0 = c \abs{\mathbf{k_0}}$. The atomic energy difference between ground and Rydberg state is $\hbar \omega_0$ and $g_0$ describes the coupling strength of the atom with the plane-wave mode.
The potential $V_{i,j}$ accounts for the Rydberg-Rydberg interaction preventing the simultaneous excitation of two atoms, it justifies $\hbar g_0 \sqrt{\braket{a^\dagger_\mathbf{k_0} a_\mathbf{k_0}}} \ll V_{i,j}$ for any combination of $i,j$, embodying the fact that the atomic cloud is smaller than the Rydberg blockade for any driving field of interest.
The interaction potential will not be of further interest in this section. 

Once a photon in mode $\mathbf{k}_0$ has interacted with the cloud of atoms in the initial state $\ket{G} =\ket{g_1,g_2,\dots,g_N}$ and has been absorbed by the medium, the time evolution can be approximated for short times by expanding the evolution operator, where $\mathcal{T}$ denotes the time ordering operator,~\cite{Wodkiewicz2006}
{
\setlength{\mathindent}{0cm}
\begin{equation}
\begin{array}{rcl}
U(\tau) &=& \mathcal{T} e^{-i/\hbar \int_{t_0}^\tau \d t' V(t')} \\
        &\approx & 1 - i g_0 \sum_j \int_{t_0}^\tau \d t' \left( \sigma^\dagger_j a_\mathbf{k_0}  \e^{i \mathbf{k_0}\cdot\mathbf{r}_j} \e^{-i(\nu_0 -\omega_0)t'} +  \sigma_j a^\dagger_\mathbf{k_0}  \e^{-i \mathbf{k_0}\cdot\mathbf{r}_j} \e^{i(\nu_0-\omega_0)t'}\right).
\end{array}
\end{equation}
}
Looking at the evolution of the ground state under the evolution of this operator reveals that the cloud will be in a time-dependent superposition of the ground and a collectively excited state
\begin{equation}
  U(\tau) \ket{G}\ket{1_\mathbf{k_0}} \approx 
   \ket{G}\ket{1_\mathbf{k_0}} -i g_0\tau  \sum_j \e^{i \mathbf{k_0}\cdot\mathbf{r_j}} \ket{g_1,g_2,\dots,r_j,\dots,g_N}\ket{0}.
\end{equation}
The collective bright state, also called swept gain ``timed'' Dicke state~\cite{Scully2007, Scully2010b},
\begin{equation}
  \ket{W} = \frac{1}{\sqrt{N}} \sum_j \e^{i \mathbf{k_0}\cdot\mathbf{r_j}} \ket{g_1,\dots,r_j,\dots,g_N},    
  \label{eq:W_state}
\end{equation}
shows the collective excitation, which is delocalized among all constituents. The atoms do not couple individually to the light field but rather in a collective manner. Evaluating 
\begin{equation}
  \bra{0}\braket{W|V(t)|G}\ket{1_\mathbf{k_0}} = \sqrt{N} \hbar g_0
\end{equation}
shows that the coupling strength to exactly the mode $\vec{k}_0$ is enhanced by a factor of $\sqrt{N}$. After excitation to the bright state the system will start to relax back to the ground state again, this decay is described by the coupling of each atom to free-space, i.e. all available photon-modes. The Hamiltonian for this process in the rotating wave approximation then reads~\cite{Wodkiewicz2006, Zubairy1997}
\begin{equation}
    W(t) = \sum_\mathbf{k} \sum_j \hbar g_\mathbf{k} \left(  \sigma^\dagger_j a_\mathbf{k}  \e^{i \mathbf{k}\cdot\mathbf{r}_j} \e^{-i(\nu_\mathbf{k} -\omega_0)t} + \sigma_j a^\dagger_\mathbf{k}  \e^{-i \mathbf{k}\cdot\mathbf{r}_j} \e^{i(\nu_\mathbf{k}-\omega_0)t}\right),
\end{equation}
where $g_\mathbf{k}$ is the coupling strength of the atom to mode $\mathbf{k}$.
This form of the Hamiltonian allows to treat the decay of every atom individually. The time-evolution for the collectively excited state will then be given by~\cite{Wodkiewicz2006, Scully2007, Scully2010b}
\begin{eqnarray}
  U_W(t) &= \mathcal{T} e^{-i/\hbar \int_{t_0}^\tau \d t' W(t')}= \prod_j U^{(j)}_W \simeq \sum_j \gamma_j^\dagger \sigma_j\\
  \gamma_j^\dagger&= \sum_\mathbf{k} \frac{g_\mathbf{k} \e^{-i\mathbf{k}\cdot\mathbf{r_j}}}{\nu_k -\omega_0 +\frac{i\Gamma}{2}} a^\dagger_\mathbf{k},
\end{eqnarray}
where $\gamma_j^\dagger$ describes the spontaneous emission process of a photon from atom $j$ into mode $\mathbf{k}$, and $\Gamma$ is the Weisskopf-Wigner spontaneous emission rate. In the last step the long time limit $\tau \rightarrow \infty$ is taken. For details of the derivation of $U^{(j)}_W$ consult~\cite{Wigner1930, Zubairy1997}. 
In contrast to spontaneous emission from a single atom, the photon field arising from the decay of the bright state $\ket{W}$, is not uniformly distributed, but highly directed~\cite{Wodkiewicz2006, Scully2008, Scully2010b}. This can be seen by further investigating the process in which the excitation in $\ket{W}$ is converted to a photon in mode $\mathbf{k}$. The outgoing photon state is given by
\begin{eqnarray}
  \bra{G}U_W\ket{W}\ket{0} &= \frac{1}{\sqrt{N}} \sum_j \e^{i \mathbf{k_0}\cdot\mathbf{r_j}} \gamma_j^\dagger \ket{0} \\
  &\approx \frac{\sqrt{N}}{V}\sum_\mathbf{k} \frac{g_\mathbf{k}(2\pi )^3}{\nu_k -\omega_0 +\frac{i\Gamma}{2}} \ket{1_\mathbf{k}} \delta^{(3)}(\mathbf{k_0} -\mathbf{k}). 
  \label{eq:directed_emission_and_enhanced_decay}
\end{eqnarray}
For the last step the atomic cloud is assumed to have a fairly high number density, such that the summation over $j$ can be calculated as an integral as follows
\begin{eqnarray}
   \sum_j \e^{i (\mathbf{k_0}-\mathbf{k})\cdot\mathbf{r_j}} &\approx & \frac{N}{V} \int_V \d^{\rs (3)} \mathbf{r} \e^{i (\mathbf{k_0}-\mathbf{k})\cdot\mathbf{r_j}} \\
   &=& \frac{N (2 \pi)^3}{V} \delta^{(3)}(\mathbf{k_0} -\mathbf{k}),
\end{eqnarray}
where $V$ is the volume of the atomic medium. The incoming photon thus dictates the direction for emission of the outgoing photon field due to the fact that all atoms inherit the incoming photon's phase relation. Furthermore, not only the excitation but also the emission process is collectively enhanced and the atom decays  $\approx N$ times faster; note that the emission rate into all other directions is still lower bounded by the single atom decay rate $\Gamma$~\cite{Scully2007}. The previous assumption of a fairly high number density implies that the sample exhibits a significant optical depth, thus the photon amplitude would exhibit an exponential decrease as it traverses the atomic cloud, and thus not all atoms would contribute equally to the bright state~\cite{Scully2007}. This means that the bright state would not be the only one being excited by the photon absorption, and other subradiant or dark states get excited~\cite{Scully2007}.  These states have also only a single Rydberg excitation and are thus also affected by the Rydberg blockade. The dark states $\{ \vert D_j \rangle \}_{j = 1}^{N-1} $, if neglecting virtual photon processes~\cite{Scully2007, Scully2009}, do not couple to the photon field and thus decay slowly compared to the $\ket{W}$ while the single atom decay still acts as a lower bound to the decay rate. The exact nature of the dark states, is of minor relevance here, and the coupling between the bright and dark states will be considered as an additional dephasing mechanism in the following. The dephasing into the dark state manifold in a cold atom experiment setting arises partly from inhomogenities of the trapping potential and thermal motion but also from virtual exchange processes~\cite{Scully2009}.   In fact, to precisely model the decay dynamics of a real superatom the coupling with subradiant states due to virtual exchange of photons must be accounted for~\cite{stiesdalObservationCollectiveDecay2020}.

\subsection{Enhanced Emission into Gaussian Mode}
\label{sec:gaussian_emission}

Having discussed the toy-model case of a single photon and a frozen gas, we will now try to connect this system to a more physical picture, and introduce a model resembling an experimental setting more closely where the atoms are trapped in an harmonic trap and coupled to a propagating Gaussian mode, see also~\cite{Hofferberth2017c}.

The atomic cloud is assumed to be a thermal, non-quantum-degenerate gas, such that it is uncorrelated on a wavelength comparable to the light's wavelength $\lambda$.  The cloud will be assumed to be described by a transverse extent $\sigma_r$, and an axial extent $\sigma_z$ along the propagation direction of the photon field. The atomic cloud can be described by the  field operators $\Psi^\dagger_g(\vec r)$, creating a ground state $\ket{g}$ atom at position $\vec r$, and  $\Psi^\dagger_r(\vec r)$, creating a Rydberg atom $\ket{r}$ at position $\vec r$. Their commutation relations are $[\Psi_\sigma(\vec{r}), \Psi^\dagger_{\sigma'}(\vec{r}')] = \delta_{\sigma \sigma'} \delta(\vec{r} - \vec{r}')$ and $[\Psi_\sigma(\vec{r}), \Psi_{\sigma'}(\vec{r}')] =[\Psi^\dagger_\sigma(\vec{r}), \Psi^\dagger_{\sigma'}(\vec{r}')]= 0 $ with $\sigma \in \{g, r \}$.

The ground state of the atomic system is then given by $\ket{G} = \sqrt{1/N!} \prod_{i = 1}^N \Psi_g^\dagger(\vec{r}_i) \ket{0}$. The atomic distribution is randomly distributed such that the averaged ground state density is
\begin{equation}
n(\vec{r}) = \langle \bra{G} \Psi_g^\dagger(\vec{r}) \Psi_g(\vec{r}) \ket{G} \rangle_\text{dis} 
= n_0 \exp\left( - \frac{z^2}{2 \sigma_z^2} - \frac{x^2 + y^2 }{2 \sigma_r^2} \right)\,  , 
\label{eq:gaussian_density}
\end{equation}
where $\langle \cdots \rangle_\text{dis}$ denotes the ensemble average over many experimental realizations. In order to describe the process of an excitation of an atom from the ground to the Rydberg state at position $\vec{r}$, we introduce the operators $S^+(\vec{r}) = \Psi_r^\dagger(\vec{r}) \Psi_g(\vec{r})$ and $S^-(\vec{r}) = \Psi_g^\dagger(\vec{r}) \Psi_r(\vec{r})$ for the transition from the Rydberg state to the ground state. The commutator of these operators satisfies
\begin{equation}
[S^-(\vec{r}) , S^+(\vec{r}')] = \delta(\vec{r} - \vec{r}') \left( \Psi_g^\dagger(\vec{r}) \Psi_g(\vec{r}) - \Psi_r^\dagger(\vec{r}) \Psi_r(\vec{r}) \right)\, . 
\end{equation}
The microscopic Hamiltonian describing the coupling of the propagating light field to the atomic cloud within the dipole and rotating-wave approximation is then described by 
\begin{equation}
H = \int \frac{\mathrm{d}\vec{q}}{(2\pi)^3}\, \hbar \omega_\vec{q} a^\dagger_\vec{q} a_\vec{q} + g_0 \int \mathrm{d} \vec{r} \left[ S^+(\vec{r}) \mathcal{E}(\vec{r}) + \mathcal{E}^\dagger(\vec{r}) S^-(\vec{r}) \right]\, . 
\end{equation}
Here, $g_0$ describes the dipole matrix element for the transition. The electric field operator reads
\begin{equation}
\mathcal{E}(\vec{r}) = \sum_\mu \int \frac{\mathrm{d} \vec{q}}{(2\pi)^3} c^\mu_\vec{q} a_{\vec{q}} e^{i \vec{q} \cdot \vec{r}} \, ,
\end{equation}
with $c^\mu_\vec{q} = i \sqrt{\omega_\vec{q} 2 \pi \hbar} \, \vec{p} \cdot \boldsymbol{\epsilon}_\vec{q}^\mu$ being the normalization and influence of the polarization $\boldsymbol{\epsilon}_\vec{q}^\mu$. In contrast to the discussion above, the incoming field is not assumed to be a plane wave but a Gaussian beam that propagates along the $z$ direction with a beam waist $w_0$ and polarization parallel to the direction of the dipole $\vec{p}$. At the focal point, the Gaussian beam has a transverse mode area $A= \pi w_0^2 / 2$. As discussed above, the state that can be addressed by the laser field is the bright state $\ket{W}$, where only a single atom is excited. In contrast to the case of an incoming plane wave, for an arbitrary incoming field with mode function $u(\vec{r})$, the bright state reads
\begin{equation}
\ket{W} = \frac{1}{\sqrt{N}} \int \mathrm{d} \vec{r} \, u(\vec{r}) S^+(\vec{r}) \ket{G}\, ,
\end{equation}
where $N$ describes the number of atom overlapping with the incoming photon field, 
\begin{equation}
N = \int \mathrm{d}\vec{r} \, \vert u(\vec{r}) \vert^2 \bra{G} \Psi_g^\dagger(\vec{r}) \Psi_g(\vec{r}) \ket{G} \, . 
\end{equation}
The exact number of atoms contributing to the overlap will depend on the specific realization, yet on average, the number fluctuations for the random distribution will be suppressed as $\Delta N / \bar{N} \propto 1 / \sqrt{\bar{N}}$, where $\bar{N}$ denotes the mean number of atoms contributing to the superatom. In an experimental realization~\cite{Hofferberth2017c}, the average number of particles was $\approx 10^4$ such that those fluctuations may be neglected.
For the Gaussian density distribution Eq.~(\ref{eq:gaussian_density}) with widths $\sigma_z$ and $\sigma_r$, the field density of the propagating Gaussian mode in the experimentally relevant regime where $\lambda \ll w_0, \sigma_z$ and $w_0 \ll \sigma_r$, can be well approximated by $\vert u(\vec{r}) \vert^2 = \exp(- 2 (x^2 + y^2)/ w_0^2)$ and the average number of particles contributing to the superatom is
\begin{equation}
\bar{N} = \int \mathrm{d} \vec{r} \, \vert u(\vec{r}) \vert^2 n(\vec{r}) = \frac{(2\pi)^{3/2}}{4} w_0^2 \sigma_z n_0 = \sqrt{2\pi} \sigma_z A n_0\, . 
\end{equation}
Next, we show that the emission rate of the state $\ket{W}$ is enhanced for the emission into the mode $u(\vec{r})$. As a first step, we determine the rate for the emission into a mode with momentum $\vec{q}$ and polarization $\mu$ within Fermi's Golden Rule, which reads

{
\setlength{\mathindent}{0cm}
\begin{equation}
\begin{array}{rcl}
\Gamma_{\vec{q}, \mu} &=& \frac{2\pi}{\hbar} \delta(\hbar \omega - \hbar c \vert \vec{q} \vert) \vert \bra{1_{\vec{q}, \mu}} \bra{G} H \ket{W} \ket{0} \vert^2 \nonumber \\
&=& \frac{2\pi}{\hbar} \delta(\hbar \omega - \hbar c \vert \vec{q} \vert) \left\vert \frac{g_0}{\sqrt{N}} \int \mathrm{d} \vec{r} \, \mathrm{d} \vec{r}' \, u(\vec{r}') \bra{1_{\vec{q}, \mu}} \bra{G} \mathcal{E}^\dagger(\vec{r}) S^-(\vec{r}) S^+(\vec{r}') \ket{G} \ket{0} \right\vert^2 \nonumber \\
&=& \frac{2\pi}{\hbar} \delta(\hbar \omega - \hbar c \vert \vec{q} \vert) \left\vert \frac{g_0}{\sqrt{N}} (c_{\vec{q}}^\mu)^* \int \mathrm{d} \vec{r} \, \mathrm{d} \vec{r}' \, u(\vec{r}')  e^{- i \vec{q} \cdot \vec{r}} \bra{G} \Psi^\dagger_g(\vec{r}) \Psi_r (\vec{r}) \Psi^\dagger_r(\vec{r}') \Psi_g(\vec{r}') \ket{G} \right\vert^2 \nonumber \\
&=& \frac{2\pi}{\hbar} \delta(\hbar \omega - \hbar c \vert \vec{q} \vert) \left\vert \frac{g_0}{\sqrt{N}} (c_{\vec{q}}^\mu)^* \int \mathrm{d} \vec{r}  \, u(\vec{r}) e^{-i \vec{q} \cdot \vec{r}} \bra{G} \Psi_g^\dagger(\vec{r}) \Psi_g(\vec{r}) \ket{G} \right\vert^2 \nonumber \\
&=& \frac{2\pi g_0^2}{\hbar} \delta(\hbar \omega - \hbar c \vert \vec{q} \vert) \vert c_\vec{q}^\mu \vert^2 \int \mathrm{d} \vec{r} \, \mathrm{d} \vec{r}' \, e^{- i \vec{q}\cdot(\vec{r} - \vec{r}')} u(\vec{r}) u^*(\vec{r}') \frac{n_g(\vec{r}) n_g(\vec{r}')}{N} \, ,  
\end{array}
\end{equation}
}

where $n_g(\vec{r}) = \bra{G} \Psi_g^\dagger(\vec{r}) \Psi_g(\vec{r}) \ket{G}$. Since the positions of the atoms are random and fluctuate within each experimental realization, we have to average this quantity leading to
{
\setlength{\mathindent}{1cm}
\begin{eqnarray}
\bar{\Gamma}_{\vec{q}, \mu} & = \langle \Gamma_{\vec{q}, \mu} \rangle_\text{dis} \\ 
&= \frac{2\pi g_0^2}{\hbar} \delta(\hbar \omega - \hbar c \vert \vec{q} \vert) \vert c_\vec{q}^\mu \vert^2 \int \mathrm{d} \vec{r} \, \mathrm{d} \vec{r}' \, e^{- i \vec{q} \cdot (\vec{r} - \vec{r}')} u(\vec{r}) u^*(\vec{r}') \left\langle \frac{n_g(\vec{r}) n_g(\vec{r}') }{N}\right\rangle \, . 
\end{eqnarray}
}
In order to calculate the correlator, we use the general property
\begin{equation}
\langle n_g(\vec{r}) n_g(\vec{r}') \rangle_\text{dis} = g^{(2)}(\vec{r}, \vec{r}') n_g(\vec{r}) n_g(\vec{r}') + n_g(\vec{r}) \delta(\vec{r} - \vec{r}') \, , 
\end{equation}
where $g^{(2)}$ is the two-body correlation function of the atoms. For a thermal gas well above quantum degeneracy and atoms that are randomly distributed within the trap, the atoms are uncorrelated on distances comparable to the optical wavelength and we can approximate $g^{(2)} = 1$ in the limit of a large number of atoms. Thus, the averaged emission rate from the state $\ket{W}$ into a mode with momentum $\vec{q}$ and polarization $\mu$ is given by
{
\setlength{\mathindent}{1cm}
\begin{eqnarray}
\bar{\Gamma}_{\vec{q}, \mu} &= \frac{2\pi g_0^2}{\hbar} \delta(\hbar \omega - \hbar c \vert \vec{q} \vert) \vert c_\vec{q}^\mu \vert^2  \left[1 + \frac{1}{\bar{N}} \left\vert \int \mathrm{d}\vec{r} \, e^{- i \vec{q} \cdot \vec{r}} u(\vec{r}) n_g(\vec{r}) \right\vert^2 + \mathcal{O}(\Delta N / \bar{N}) \right] \, . 
\label{eq:emission_rate_averaged}
\end{eqnarray}
}

The first term gives rise to the standard spontaneous decay rate $\Gamma = 4 g_0^2 \omega^3/(3 \hbar c^3)$ of every individual atom which is randomly directed into any mode $\vec q$, this contributes an incoherent process. The second term characterizes the collective enhancement of the decay into a specific mode, corresponding to a scenario as described in~\cite{Hofferberth2017c}, where the wavelength $\lambda$ is much shorter than the transverse beam waist $w_0$, or the extent of the cloud. Then, the collective decay only provides significant contributions for momenta in the forward direction with an angle

\begin{equation}
    \sin^2\theta \leq \frac{\lambda^2}{\pi w_0^2},
\end{equation}
corresponding to the opening angle of a Gaussian beam. This result suggests to calculate the emission rate into the forward propagating mode $u(\vec{r})$, which we might approximate as $u(\vec{r}) \approx \exp(- (x^2 + y^2) / w_0^2)\exp(i k z)$ for the experimentally relevant parameter regime. In addition, we assume this mode to have a polarization aligned with the atomic dipole $\vec{p}$. This leads to
\begin{eqnarray}
\kappa &= &\frac{2\pi g_0^2}{\hbar} \int \frac{\mathrm{d}k}{2\pi} \delta (\hbar \omega - \hbar c k) \frac{ \vert c_k^\mu \vert^2}{A} \left[1 + \frac{1}{\bar{N}} \left\vert \int \mathrm{d}\mathbf{r} \vert u(\vec{r}) \vert^2 n_g(\vec{r})\right\vert^2 \right] \\
&=& \frac{2\pi (\bar{N} + 1)g_0^2}{A} \frac{\omega}{\hbar c} \, . 
\end{eqnarray}
In the following, we can approximate $\bar{N} + 1 \approx \bar{N}$. For a transverse width of the atomic distribution $\sigma_r \leqslant w_0$, transitions into higher Gaussian modes are possible. Furthermore, the emission rate into the backward propagating Gaussian mode with momentum $-k$ reduces to the incoherent contribution as the influence of the second term in the bracket of Eq.~(\ref{eq:emission_rate_averaged}) is exponentially suppressed by $\exp( -8 \pi^2 \sigma_z^2 / \lambda^2)$ for an atomic distribution that is smooth compared to the wavelength of the incoming probe beam. 

The coupling strength of the photon field and superatom can thus be written as 
\begin{equation}
 g_{\rs{ col}}= 2 \sqrt{\kappa} = \sqrt{\frac{8 \pi \bar{N} g_0^2 \omega}{ A \hbar c}} = \sqrt{\frac{ 3 \bar{N} \: \Gamma \: \lambda^2}{2 \pi A}}.
\end{equation}
This coupling strength shows that a reduction to a two-level system is possible even in the case where the photon field is a Gaussian beam and not a more idealized plane wave. For the experimental parameters used in \cite{Hofferberth2017c}, and accounting for the change in coupling constant due to adiabatic elimination by $g_{\rs{ col}}^{\rs{eff}} = g_{\rs{ col}} \Omega_c/2\Delta$ an effective collective coupling constant in the order of $g_{\rs{ col}}^{\rs{eff}}\approx	\SI{1}{\micro\second\tothe{-1}\per\sqrt{\micro\second}}$ was obtained. To put this in terms of an effective angular Rabi frequency the rate of photons impinging on the atoms $\mathcal{R}$ (in units of $\si{\per\micro\second}$) should be considered to calculate $\Omega_{\rs{col}}^{\rs{eff}}=g_{\rs{ col}}^{\rs{eff}}\sqrt{\mathcal{R}}$. In the previous discussion, we assumed a constant number of atoms interacting with the Gaussian mode. However, in some situations, this number may vary with time, leading to a time-varying effective coupling strength~\cite{dingSizereductionRydbergCollective2022}.

In the following, we will investigate the effects of strongly coupling to a few-photon propagating light field.
\section{Free-space QED with Rydberg superatoms}
\label{sec:freeQED}
One important aspect of the collectively enhanced coupling to one forward propagating mode is that the effects of atom-light interaction can already be observed on the single- to few-photon level.

\subsection{Derivation of the master equation}

In order to derive the master equation that describes the dynamics of the superatom alone, we start with the Hamiltonian of a single two-level system coupled to a quantized light field in one dimension given by
\begin{equation}
H = \int \frac{dk}{2 \pi} \hbar c k \, a_k^\dagger a_k + \hbar \sqrt{\kappa} \left( E^\dagger(0) \sigma_{GW} + E(0) \sigma_{GW}^\dagger \right)\, . 
\label{eq:hamiltonian_free_space}
\end{equation}
In this expression, $a_k^{(\dagger)}$ denotes the annihilation (creation) operators of the photonic field, $\sqrt{\kappa} = g_\text{col}/2$ is the collectively enhanced coupling strength to the forward propagating mode, $E(x) = \left[ \sqrt{c} / (2\pi) \right] \int dk \, e^{ikx} a_k$ is the electric field operator measured in $\sqrt{\text{photons/time}}$, and $\sigma_{\alpha \beta} = \vert \alpha \rangle \langle \beta \vert$. In Equation~(\ref{eq:hamiltonian_free_space}) the dynamics of an ensemble of atoms is modeled as a single point-like, two-level system for which the collectively enhanced emission into a preferred mode is captured by the parameter $\kappa$. The derivation of the master equation uses methods described in standard textbooks~\cite{Zoller2004} and used in publications on atom-light coupling in one-dimensional waveguides (see e.g.~\cite{Zoller2015, Cirac2015}). When a probe photon has passed through the system, it irreversibly leaves such that we can solve and trace out the time dependence of the photonic part, leaving us with the effective dynamics of the atomic degrees of freedom. 

The Heisenberg equations of motion for the photonic field operators are
\begin{equation}
\partial_t a_k(t) = - \frac{i}{\hbar} [ a_k(t), H] = - i c k \, a_k(t) - i \sqrt{\kappa \, c} \sigma_{GW}(t) \, . 
\end{equation}
This equation can be formally integrated and connects the outgoing electric field to the operator $\sigma_{GW}$ that describes the coherence of the superatom:
\begin{equation}
a_k(t) = e^{- i c k (t-t_0)} a_k(t_0) - i \sqrt{\kappa \, c} \int_{t_0}^t ds \, e^{i c k (s-t)} \sigma_{GW}(s) \, .  
\label{eq:ak_solution}
\end{equation}
Here, $t_0$ denotes the initial time for which the incoming photon field has not yet reached the superatom. In the following, we can assume $t_0 = 0$ without the loss of generality. Equation~(\ref{eq:ak_solution}) leads to the useful relation between the outgoing electric field and the dynamics of the coherence of the superatom, 
\begin{eqnarray}
E(x,t) &=& E_0(ct - x) - i \sqrt{\kappa} c \int_0^t ds \, \int \frac{dk}{2\pi} e^{- i c k (t - s) + i k x} \sigma_{GW}(s) \nonumber \\
&=& E_0(ct - x) - i \sqrt{\kappa} \, \sigma_{GW}(t - x/c) \theta(x) \theta(ct - x) \, . 
\label{eq:e-field}
\end{eqnarray}
Here, $E_0$ denotes the non-interacting electric field operator and $\theta(x)$ is the Heaviside function with the definition that $\theta(0) = 1/2$. If the incoming photon field is in a coherent state, the non-interacting electric field operator may be replaced by its expectation value $\alpha(t) \equiv \langle E_0(ct) \rangle$, which characterizes the incoming photon rate $\mathcal{R}_{\mathrm{in}} = \vert \alpha(t) \vert^2$. 

The Heisenberg equation of motion for an arbitrary operator $O$ that acts on the atomic degrees of freedom only is
\begin{eqnarray}
&\partial_t O(t) = - i \sqrt{\kappa} [ O(t),  \alpha (t) \sigma^\dagger_{GW}(t) + \alpha^*(t) \sigma_{GW}(t)]  \nonumber \\
 & - \frac{\kappa}{2}\left( [ O(t), \sigma^\dagger_{GW}(t)]\sigma_{GW}(t) - \sigma^\dagger_{GW}(t) [O(t), \sigma_{GW}(t)] \right) \, . 
\end{eqnarray}
Using the relation $\partial_t \langle O(t) \rangle = \Tr\{ O \partial_t \rho(t) \} $ with the reduced density matrix of the atomic system $\rho$, the dynamics of $\rho(t)$ are described by the master equation
\begin{equation}
\partial_t \rho(t) = - \frac{i}{\hbar} [ H_0(t), \rho(t)] + \kappa \mathcal{D}[\sigma_{GW}] \rho(t) \, , 
\label{eq:master_equation1}
\end{equation}
where $H_0(t) = \hbar \sqrt{\kappa} \left( \alpha^*(t) \sigma_{GW}(t) + \alpha(t) \sigma^\dagger_{GW} \right)$ describes the coherent evolution under the driving of a (classical) light field originating from the non-interacting part of the electric field operator and the Lindblad dissipator $\mathcal{D}[\sigma] \rho = \sigma \rho \sigma^\dagger -\frac{1}{2}( \sigma^\dagger \sigma \rho + \rho \sigma^\dagger \sigma)$ accounts for the dissipation due to the collectively enhanced spontaneous emission into the forward direction with rate $\kappa$. The same result can be obtained by applying the Mollow transformation to the Hamiltonian from Equation~(\ref{eq:hamiltonian_free_space})~\cite{Mollow1975} and deriving the Heisenberg equations of motion for the transformed Hamiltonian.  

Until now, only the enhanced spontaneous emission into the forward propagating mode has been taken into account while the superatom can also decay into transverse photonic modes (cf. Sec. \ref{sec:collective_emission}). The decay rate of this process is still given by the standard single-atom emission rate $\Gamma$ in free space. Furthermore, the bright state $\vert W \rangle$ will dephase into the manifold of dark states $\{ \vert D_j \rangle\}_{j = 1}^{N-1} $ due to Doppler shifts of the atoms as well as inhomogeneous shifts of the Rydberg state level and residual interactions between individual atoms mediated by virtual photons (resonant dipole-dipole interactions)~\cite{Zoller2015, Scully2008, Lehmberg1970, Chang2015b, Cirac2008}. The dark states decay into the ground state by incoherent emission of photons with rate $\Gamma$.
The incoherent effects discussed above are included into the master equation (\ref{eq:master_equation1}) phenomenologically, leading to
{
\setlength{\mathindent}{0cm}
\begin{equation}
\partial_t \rho(t) = - \frac{i}{\hbar} [ H_0(t), \rho(t)] + (\kappa + \Gamma) \mathcal{D}[\sigma_{GW}] \rho(t)  
+ \gamma_D \mathcal{D}[\sigma_{WD}] \rho(t) 
+ \Gamma \mathcal{D}[\sigma_{GD}] \rho(t) \, ,
\label{eq:master_equation2} 
\end{equation}
}
where $\gamma_D$ denotes the loss rate associated with the dephasing of the state $\ket{W}$ into the manifold of dark states $\{\ket{D_j}\}_{j = 1}^{N-1} $, cf. Figure~\ref{fig:system}(c). Since the specifics of the dark state is irrelevant, some 'dummy' dark state $\ket{D}$ representing the whole manifold is chosen. Note that here the dephasing is treated like a decay mechanism which reproduces the experimental results very well (see below). However, a more detailed analysis of the influence of the resonant dipole-dipole interactions for an atomic cloud coupled to a propagating light mode in one dimension shows that this gives rise to non-trivial dynamics~\cite{Buechler2018}.

The master equation (\ref{eq:master_equation2}) can now be used to simulate the dynamics of the coupled system of photons and the superatom. Together with the relation between outgoing and incoming field (\ref{eq:e-field}), it provides the basis for calculating expectation values and correlation functions. Defining the retarded time $s = t - x/c$, the outgoing photon flux can be computed as
\begin{eqnarray}
\langle E^\dagger(s) E(s) \rangle &= \vert \alpha(s) \vert^2 + \kappa \langle \sigma^\dagger_{GW}(s) \sigma_{GW}(s) \rangle \nonumber \\
&- i \sqrt{\kappa} \left[\alpha^*(s) \langle \sigma_{GW}(s) \rangle - \alpha(s) \langle \sigma^\dagger_{GW}(s) \rangle \right] \, . 
\end{eqnarray}

In the superatom setup, the strong directional emission into the same mode that was used for excitation provides a natural waveguide system without the need for any additional structures confining the light field resulting in so-called "waveguide QED without a waveguide" \cite{sheremetWaveguideQuantumElectrodynamics2023}. 
This can also be quantified using the $\beta$-factor which specifies the emission rate into the waveguide compared to the total emission rate, $\beta = \kappa/ (\kappa + \Gamma)$. In a recent experiment~\cite{Hofferberth2017c}, $\beta = 0.86$ was reported. However, this definition of the $\beta$-factor does not take into account the non-radiative contributions from the dephasing into the dark states which we can also interpret as an emission that is not into the waveguide. Defining $\beta_\text{coh} = \kappa / (\kappa + \gamma_D + \Gamma)$ as a measure of the ratio between the emission into the waveguide and the total losses, the experimental parameters result in $\beta_\text{coh} \approx 0.23$.

\subsection{Dynamical phase diagram}

\begin{figure}
\centering
\includegraphics[width=\textwidth]{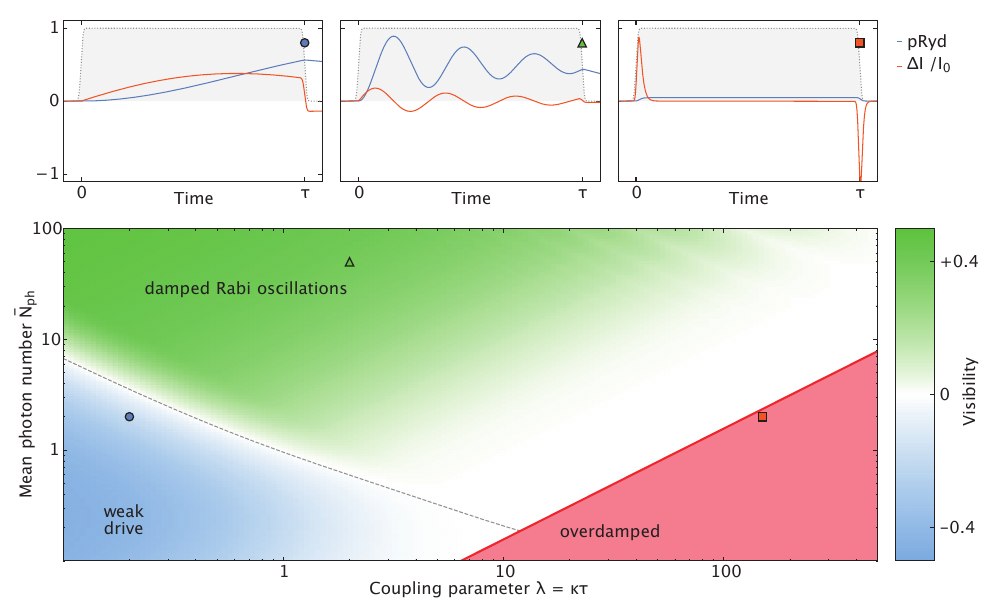}
\caption{Dynamical phase diagram of a driven atom in free space. (Bottom panel) The phase diagram shows the visibility of Rabi oscillations in the outgoing field, defined as $\text{max}_{0 \leq t \leq \tau} [ \rho_{WW} (t) ] - \rho_{WW}( t = \infty)$, of an ideal ($\Gamma = \gamma_D = 0$) two-level system driven by a propagating light field. In contrast to cavity QED, the coupling of the photons to the atom and their decay are not independent in free-space and waveguide QED. For a large coupling to the propagating mode ($\lambda = \kappa \tau \gg 1$), the enhanced emission into this mode results in an overdamped system, where the number of photons required to observe Rabi oscillations increases with the coupling strength. For $\lambda \ll 1$, a large number of photons is required to drive the system with a $\pi$-pulse, defining a crossover (dashed line) between the regime of damped Rabi oscillations and the weak driving regime at lower mean photon number. (Top panels) Examples of the variation of the Rydberg population ($p_\text{Ryd}$, blue) and the relative modulation of the intensity of the driving field, ($\Delta I /  I_0 = (I_\text{in} - I_\text{out})/I_\text{in}$, orange). The pulse shape of the incoming driving field is shown in shaded gray. }
\label{fig:phase_diagram}
\end{figure}

The essential physics of the single superatom interacting with the propagating light mode is best understood when neglecting all internal dephasing mechanisms and incoherent spontaneous emission processes. Then, the dynamics of the system are given in terms of the master equation (\ref{eq:master_equation1}). It is important to note that the driving strength and the decay are not independent of each other and increasing the coupling of the photons to the superatom inevitably leads to an increased emission rate in this model. Note that this is in contrast to cavity QED where the coupling of the emitter to the light field can be tuned independently from its decay and thus a genuine strong coupling regime with Rabi oscillations exists. In the case of a constant input pulse of some length $\tau$ with a constant photon rate $ \vert \alpha \vert^2$, the master equation can be solved analytically and the occupation of the Rydberg level, given by the component $\rho_{22}$ of the density matrix, is

\begin{equation}
\rho_{22}(t) = \frac{4 \kappa \vert \alpha \vert^2}{\kappa^2 + 8 \kappa \vert \alpha \vert^2} \left[1 - \left( \frac{3 \kappa}{4 \Omega_\text{eff}} \sin( \Omega_\text{eff} t  ) + \cos( \Omega_\text{eff} t )\right)e^{- \frac{3\kappa t}{4}} \right]
\end{equation}

\noindent
with the effective Rabi frequency $\Omega_\text{eff} = \sqrt{4 \kappa \vert \alpha \vert^2 - \left(\frac{\kappa}{4} \right)^2}$. The interplay between the driving strength and decay is then best studied by introducing the dimensionless coupling constant 
\begin{equation}
\lambda = \kappa \tau
\end{equation}
and the mean photon number per pulse
\begin{equation}
\bar{N} = \int dt \, \vert \alpha(t) \vert^2 \, ,
\end{equation}
and the physics can be cast into the dynamical phase diagram shown in Figure \ref{fig:phase_diagram}, where the visibility of the Rabi oscillations is plotted.
One can distinguish three different regimes: the overdamped regime, where the decay exceeds the driving and no Rabi oscillations occur, the weak driving regime, where the driving exceeds the decay but is not strong enough to drive Rabi oscillations, and finally the regime in between, where it is possible to observe intrinsically damped Rabi oscillations.. The boundary between the intermediate regime, where Rabi oscillations are visible and the overdamped regime is defined by the critical coupling strength for which the effective Rabi frequency becomes imaginary. In terms of the dimensionless quantities $\lambda$ and $\bar{N}$, this occurs for 
\begin{equation}
\lambda > \frac{\bar{N}}{64} \, .  
\end{equation}
In this regime, every photon excites the atom but the atom also decays again very quickly. The outgoing photon wave packet thus has correlations on a very short length scale $\sim c \tau$, while they vanish on larger distances. Consequently, no Rabi oscillations are visible but there is a finite probability for the atom to be excited.
On the other hand, there is a crossover from the weak drive regime to the regime of Rabi oscillations, that can be defined by the condition that $\Omega_\text{eff} \tau = \pi$. In the limit of weak coupling, $\lambda \ll 1$, this gives a critical photon number
\begin{equation}
\bar{N}_c = \frac{\pi^2}{4 \lambda} \, . 
\end{equation}
The regime $\lambda \ll 1$ but $\bar{N} \gg \lambda$ corresponds to the classical limit, where each photon has only a very weak coupling to the atom and the wave function of the photons is not changed. As a consequence, the outgoing photons are still described by a product state and no correlations between atom and photons occur.
In between those two regimes, where $\lambda \sim 1$, correlations between the photons and the superatom occur and Rabi oscillations can be observed also for a relatively low number of photons. Since the atom can only absorb a photon if it is not excited, it will mediate an effective interaction between the photons. 
Note that in the case of spontaneous emission into other modes and additional dephasing of the $\vert W \rangle$-state ($\Gamma, \, \gamma_D \neq 0$), the overdamped regime also occurs for weak coupling strength ($\lambda \ll 1$) and low mean photon number ($\bar{N} \ll 1$).

\subsection{Two-photon and three-photon correlations}
As mentioned above, one interesting aspect of the strong optical nonlinearity provided by the superatom is the presence of an effective photon-photon interaction. The influence of this interaction can be determined by studying the correlations imprinted onto initially uncorrelated photons, which are in general quantified by the $n$-body correlation function~\cite{glauberQuantumTheoryOptical1963}
\begin{equation}
g^{(n)}(s_1, \ldots, s_n) = \frac{\langle E^\dagger(s_1) \cdots E^\dagger(s_n) E(s_n) \cdots E(s_1) \rangle}{\langle E^\dagger(s_1) E(s_1) \rangle \cdots \langle E^\dagger(s_n) E(s_n) \rangle} \, . 
\end{equation}

Unlike the case of the intensity, where the calculation of equal-time correlations was sufficient, calculating for example two-point correlation functions involves the determination of unequal-time correlations.

Within the master equation formalism, it is possible to determine unequal-time correlations using the quantum regression theorem~\cite{Lax1968}, which in general relies on the Born-Markov approximation. However, in our system of interest, where the emitted photons never interact with the atom again, the quantum regression theorem is indeed exact. This also extends to the case of multiple superatoms in an array as long as retardation effects between the superatoms can be neglected. 

\begin{figure}
\centering
\includegraphics[width=0.5\textwidth]{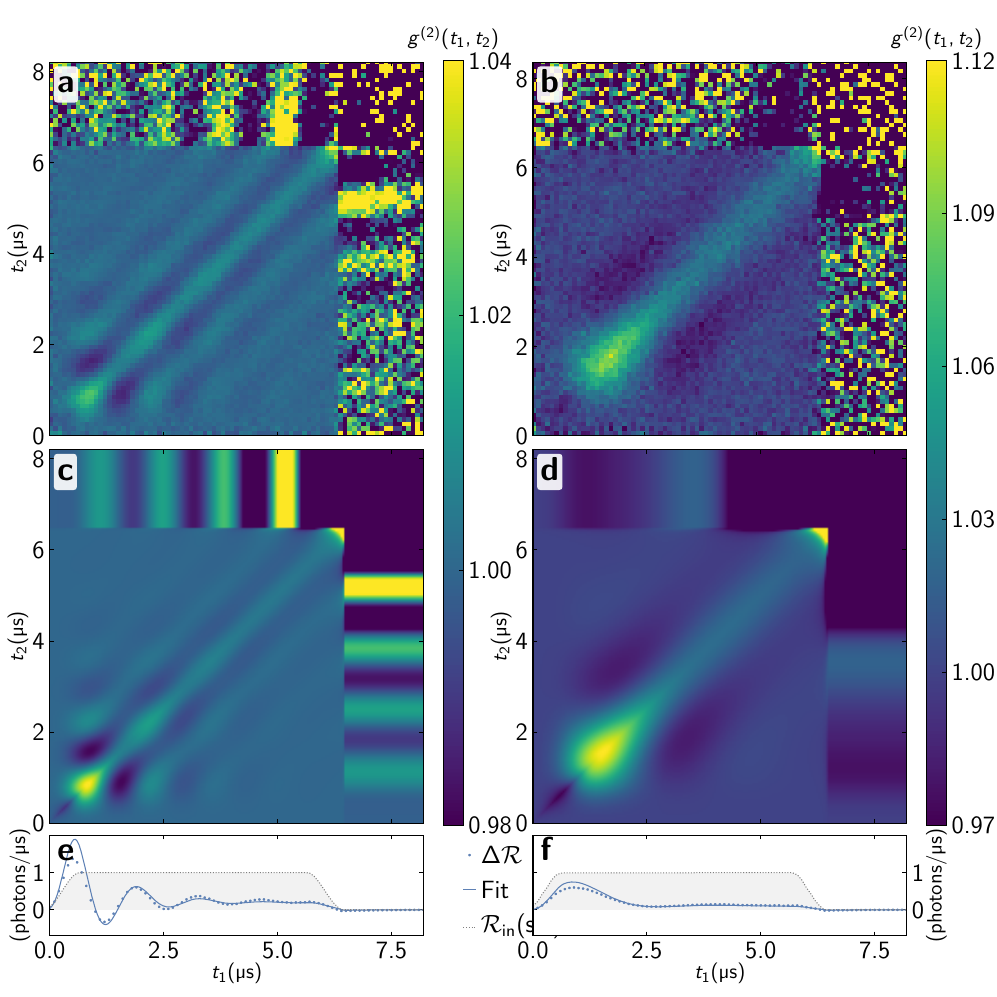}
\caption{Second order correlations of the outgoing photon field. (a) and (b) show the experimentally measured two-photon correlations $g^{(2)}(t_1,t_2)$ for pulses with a photon rates $\mathcal{R}_\text{in} = 12.4 \, \mu \text{s}^{-1}$ and $\mathcal{R}_\text{in} = 2.6 \, \mu \text{s}^{-1}$. (c) and (d) show the calculated correlation functions using the master equation and the quantum regression theorem corresponding to (a) and (b).  (e) and (f) show the photon-rate difference observed in the photon field after traversing the superatom, the gray shaded curve shows the scaled input photon pulse shape. Figure adapted from~\cite{Hofferberth2017c}}
\label{fig:correlations_2nd_order}
\end{figure}

\begin{figure}
\centering
\includegraphics[width=0.5\textwidth]{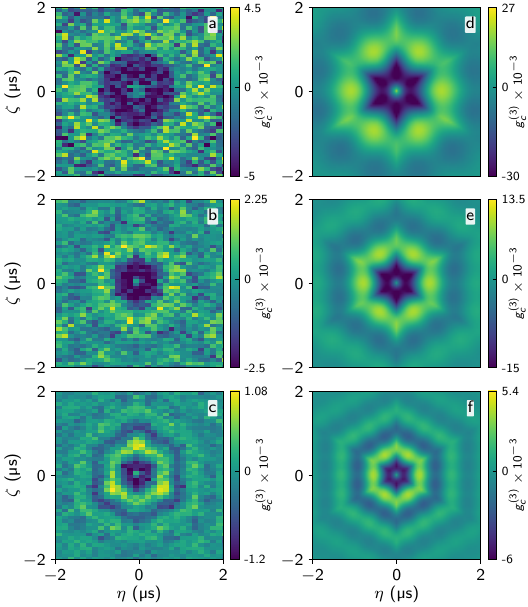}
\caption{Connected part of the three-photon correlation function.  $g^{(3)}_c$, plotted in Jacobi coordinates: $R = (s_1 + s_2 + s_3)/\sqrt{3}$, $\eta = (s_1 - s_2) / \sqrt{2}$ and $ \zeta = \sqrt{2/3} [(s_1 + s_2) / 2 - s_3]$. The experimental results in (a-c) are obtained for input rates $\mathcal{R}_\text{in} = 3.4 \, \mu \text{s}^{-1}$ (a), $\mathcal{R}_\text{in} = 6.7 \, \mu \text{s}^{-1}$ (b) and $\mathcal{R}_\text{in} = 15.2 \, \mu \text{s}^{-1}$  (c). The right row shows the corresponding simulations obtained via the master equation and quantum regression theorem.}
\label{fig:correlations_3rd_order}
\end{figure}

Figures \ref{fig:correlations_2nd_order} and \ref{fig:correlations_3rd_order} show the experimentally measured and numerically calculated two- and three-photon correlation functions for various input photon rates. One can see that there is a very good qualitative and even quantitative agreement between experiment and simulations. From the two-photon correlation function, it can be seen that two photons that are separated by a few microseconds can become entangled due to an effective photon-photon interaction mediated by the superatom. Consider therefore two photons, the first photon passes the atom and can either be absorbed or not absorbed by the atom resulting in a coherent superposition of the two cases. Another photon that passes the atom at a later time can only be absorbed and excite the atom if the first atom has not been excited which results in a spatial entanglement of the two photons. Correlations beyond the duration of the pulse originate from the collective spontaneous emission of single photons after the input pulse has left the sample which can only occur if the superatom was in the $\vert W \rangle $ state at the end of the driving pulse. 

The effective photon-photon interaction which is mediated by the Rydberg superatom also gives rise to non-trivial three-photon correlations~\cite{Hofferberth2018} as is shown in Figure \ref{fig:correlations_3rd_order}. In order to distinguish genuine three-body correlations from contributions due to two-body correlations, it is necessary to subtract these trivial contributions via the cumulant expansion. This leads to the definition of the connected part of the three-body correlation function~\cite{kuboGeneralizedCumulantExpansion1962}:
\begin{equation}
g_c^{(3)}(s_1,s_2,s_3) = 2 + g^{(3)} (s_1, s_2, s_3) - \sum_{i < j} g^{(2)}(s_i, s_j) \, .
\end{equation}
An important property of $g^{(3)}_c$ is that it vanishes if one photon is separated from the other two. While the simple single-emitter model captures the physics of the superatom-light interaction and the effective photon-photon interaction mediated through the superatom very well, a microscopic and qualitative understanding of the origin of these correlations can be gained by studying an idealized setup which ignores the additional dephasing and spontaneous emission of the excited state. It turns out that the Hamiltonian from Equation~(\ref{eq:hamiltonian_free_space}) is exactly solvable by means of the Bethe ansatz~\cite{Yudson1984, Yudson1985, Reineker2008} and the eigenstates for three photons can be characterized as a three-photon bound state, a combination of a two-photon bound state and an additional scattering photon and pure scattering states. While the three-photon bound state provides the dominant contribution to the three-photon bunching signal in Figure \ref{fig:correlations_3rd_order}, the contributions from the scattering states and the two-photon bound states are still significant. It has been suggested that, by including an additional level into the dynamics using a driving field, the effective photon-photon interaction can be tailored from attractive to repulsive~\cite{iversenSelforderingIndividualPhotons2022}.

\section{Quantum optics applications of Rydberg superatoms}

The strong matter-light coupling achievable with Rydberg superatoms opens the possibility to manipulate light at the single-photon level. Two applications that have been demonstrated so far are the on-demand generation~\cite{Pfau2018b,Kuzmich2012b,hellerRamanStorageQuasideterministic2022} and the deterministic subtraction~\cite{Hofferberth2016e} of single photons.


Single-photon sources have been implemented either by probabilistically generating photon pairs by parametric down conversion (PDC) and using one of the photons in the pair to signal the presence of the other one, or by retrieving a single excitation stored in an atomic, molecular or solid-state medium~\cite{Polyakov2011} for which achieving stable frequency and narrow linewidth continues to be a challenge. The need for scalability has shifted the focus to the latter types of sources with which a photon can be obtained on demand. While some of these sources are based on single emitters such as individual atoms, ions, molecules, quantum dots or color centers in diamonds where the limited available states of the system constrain it to host only a single quantum, others rely on atomic ensembles in which single-photon emission is obtained by using very weak pulses of light to initially store the excitation in the system~\cite{Zoller2001} making them a semi-probabilistic source since the storage time is limited by atomic motion. Nonetheless, single-emitter based photon sources usually require the use of some kind of optical cavity to coerce the emission into the preferred mode while alternative schemes suggest the use of precisely controlled adiabatic passage sequences to transfer an excitation from a single emitter to an extended atomic ensemble~\cite{petrosyanDeterministicFreeSpaceSource2018a}. By implementing single-photon sources with Rydberg superatoms these shortcomings can be overcome.

\begin{figure}
\centering
\includegraphics{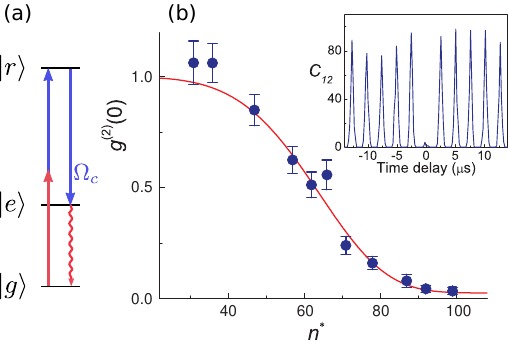}
\caption{\textbf{Single-photon source.} (a) An excitation is initially stored by driving the two-photon transition $\ket{g}\rightarrow\ket{e}\rightarrow\ket{r}$. The stored photon is retrieved by turning on the control field $\Omega_c$ (b) Experimental results adapted from~\cite{Kuzmich2012b}. Measured second-order intensity correlation function at zero time delay $g^{(2)}(0)$ as a function of the effective principal quantum number $n^*=n-\delta_S$ of the upper-level $\ket{r}$, with $\delta_S$ the quantum defect. (Inset) Cross-correlated coincidence counts $C_{12}$ as a function of time delay for $\ket{r}=\ket{102S_{1/2}}$. As $n^*$ increases the ensemble becomes fully blocked limiting the number of storable excitations to one.}
\label{fig:photon_source}
\end{figure}

Rydberg-superatom based photon sources operate by exploiting the Rydberg blockade to deterministically store one, and only one, excitation in an atomic ensemble and retrieving the excitation as a photon when needed. The excitation is stored by driving the two-photon transition as shown in Figure~\ref{fig:photon_source}a. The key factor for the operation of the single-photon source is that the extent of the atomic cloud fits within the blockade volume so that only a single excitation is stored. This excitation is stored as the collective superposition of states $\ket{W}$ as in Equation~(\ref{eq:W_state}) which contains the mode information of the storage fields. Therefore, when the photon is retrieved by driving the $\ket{r}\rightarrow\ket{e}$ transition it triggers a collective spontaneous emission process on the $\ket{e}\rightarrow\ket{g}$ transition into the same optical mode as the storage light. Moreover, the collective nature of the emission results in an enhanced decay rate, further localizing the emitted photon  temporally~\cite{Hofferberth2017c}. The frequency of the emitted photons is determined by the atomic transitions and therefore has a precise and stable value in contrast to solid-state based sources.

The principle of a Rydberg-superatom based single-photon source was first demonstrated by Dudin and Kuzmich in~\cite{Kuzmich2012b} using an ensemble of cold ($\sim\SI{10}{\micro\kelvin}$) ${}^{87}\mathrm{Rb}$ atoms constrained to be smaller than the blockade volume by using a tightly focused excitation beam ($w_e=\SI{9}{\micro\meter}$) as well as a 1D optical lattice to confine the cloud longitudinally ($w_l=\SI{15}{\micro\meter}$). As shown in Figure~\ref{fig:photon_source} they observed high-quality single-photon statistics when the principal quantum number of the Rydberg state used $n> 90$, that is, when the Rydberg interaction is large enough to fully block the entire ensemble.  More recently, Ripka et. al. demonstrated~\cite{Pfau2018b} a single-photon source based on the same principle but using a room-temperature atomic vapor as opposed to a cold atomic cloud, dramatically reducing the complexity of this approach. In their experiment the vapor was contained in a wedge-shaped glass micro cell. Again, the excitation beam was tightly focused ($w_e=\SI{1.45}{\micro\meter}$) and by translating the cell perpendicularly to the excitation beam the longitudinal thickness of the addressed atomic ensemble was controlled. In this manner, when reducing the size of the cloud enough, single-photon statistics were also demonstrated. An important difference between these two cases is the time scale in which they operate although in both, the essential processes are well described by the model presented in sections~\ref{sec:superatom} and~\ref{sec:freeQED}.

After the demonstration of Rydberg-based single-photon sources, further research was conducted to probe the indistinguishability of the emitted photons~\cite{ornelas-huertaOndemandIndistinguishableSingle2020,padron-britoProbingIndistinguishabilitySingle2021} and even to implement a quantum CNOT gate based on a single-photon source and linear optical elements~\cite{shiHighfidelityPhotonicQuantum2022}. 



\begin{figure}
\centering
\includegraphics{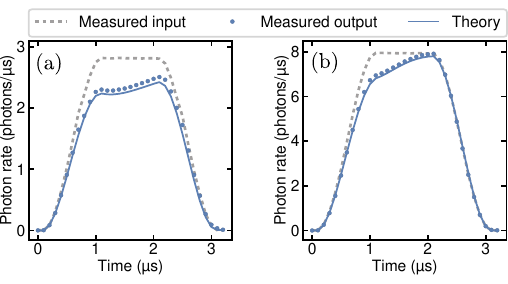}
\includegraphics{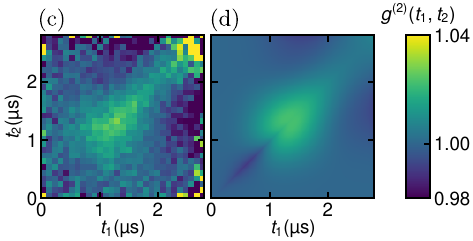}
\caption{\textbf{Single-photon absorber.} Comparison of the theory presented in section~\ref{sec:freeQED} with experimental measurements of~\cite{Hofferberth2016e}. (top) Measured (points) and simulated (solid lines) pulse shapes for 5.65 (a) and 15.76 (b) average input photons. The change of the medium from opaque to transparent after the absorption of the first photon leads to a change in the transmitted pulse shape compared to the pulse sent into the medium (dashed). (bottom) Measured (c) and simulated (d) correlation functions of the outgoing light. The photon subtraction operation results in the appearance of a bunching feature on the correlations.}
\label{fig:photon_absorber}
\end{figure}

Another important application of Rydberg superatoms is the manipulation of light fields. In particular, taking advantage of the superatom as a saturable absorber it can be used to deterministically remove individual photons from a light pulse. However, a single two-level system is not enough to act as a single-photon absorber since the process of absorption by a ground state atom is just as likely as stimulated emission by an excited one when a pulse of light passes through. Therefore, a mechanism is required to transfer the absorbed excitation to some other state that does not radiate. As discussed in section~\ref{sec:collective_emission}, the collection of dark states $\{ \vert D_j \rangle_{j = 1}^{N-1} \}$ is not coupled to the light but still contains a Rydberg excitation that can block the medium. By engineering fast dephasing $\gamma_D$ from $\ket{W}$ into these dark states, an absorbed photon is converted to a stationary Rydberg excitation preventing stimulated emission caused by the propagating light mode~\cite{Buechler2011}.

The operation of this single photon absorber has been demonstrated by Tresp et. al. in~\cite{Hofferberth2016e}. In this work an atomic cloud (${}^{87}\mathrm{Rb}$ at \SI{8}{\micro\kelvin}) is confined to a small volume ($\sigma_r=\SI{10}{\micro\meter}$, $\sigma_z=\SI{6}{\micro\meter}$) using a three-beam optical dipole trap. The Rydberg state used to ensure a full blockade of the cloud was $\ket{121S_{1/2},m_J=1/2}$ for which the blockade radius is $r_B\approx\SI{17}{\micro\meter}$. In Figure~\ref{fig:photon_absorber} the experimental results are shown, including the shape of the pulses going into and out of the superatom as well as the resulting second-order correlation function of the outgoing light. In this figure the experimental results are compared with the behaviour simulated with the theory presented in section~\ref{sec:freeQED}.

When sending coherent pulses through a single photon absorber their photon statistics will be altered. While the number of photons on the incoming pulses will follow a Poisson distribution for which the width of the distribution is proportional to the square root of its mean, for the outgoing pulses this will not be the case. The outgoing pulses will have a photon missing while the width of the distribution will stay the same as for the input statistics which leads to a super-Poissonian behaviour. This contrast with the fact that applying the anihilation operator on a coherent state leaves it unchaged. This can be understood because the anihilation operator $a=\sum_{n=1}^\infty \sqrt{n}\ket{n-1}\bra{n}$ has a different amplitude for removing a photon depending on the photon number while the photon subtraction operation $s=\sum_{n=1}^\infty\ket{n-1}\bra{n}$ removes a photon indistinctly for every photon number state. This fact results in bunching features appearing in the correlation function.

Additionally to measuring the photon statistics, it is possible to detect the photon subtraction by other means. In this system the subtracted photon remains in the atomic cloud as a Rydberg excitation. As the Rydberg electron is only loosely bound to the atom it can be ionized using moderate electric fields and detected with high probability. Then, by measuring the ion statistics, the fact that at most a single photon has been removed from the incoming light pulse can be verified. Furthermore, a linear array of superatoms can be created to realize number-resolved photon detection as shown by~\cite{stiesdalControlledMultiphotonSubtraction2021}. As the light passes through the array and is absorbed, each photon is converted to a Rydberg excitation which, by ionizing these atoms, the number of absorbed photons can be obtained by counting the detected ions. With advanced optical tweezers,  the approach could be generalized to absorb or count large numbers of photons by creating on-demand arrays of superatoms~\cite{wangPreparationHundredsMicroscopic2020}.

\section{Summary}
This tutorial provided a short framework to illustrate the necessary ingredients required to render a cloud of individual atoms to function as a collective excitation - the Rydberg superatom. 
Further it gave an instructive overview on how to tackle the problem of a single emitter coupled to a propagating mode. The key features of the superatom, the directed emission and the collectively enhanced coupling strength, were derived from simple assumptions and eventually cast into the form of a master equation describing the coupling of a propagating photon field and a single superatom. 
One important feature of superatoms coupled to propagating photon fields is the integrability of the system, due to the absence of backaction from the photonic mode; this also holds for cascaded systems of multiple superatoms. The discussed experiments implement key ingredients towards more complex quantum networks.

Rydberg superatoms created from free-space ensembles of ultracold atoms are an unexpected implementation of waveguide QED~\cite{sheremetWaveguideQuantumElectrodynamics2023}.
Due to the collective nature of the excitation, it is possible to reach comparatively high coupling efficiencies between Rydberg superatoms and propagating light fields and thereby enable manipulation of single photons.

This enhanced atom-light coupling has increased attention on Rydberg superatoms for purposes within quantum computation. Thus Rydberg superatoms offers the option of fast preparation, operation, and readout of Qubits \cite{xuFastPreparationDetection2021,spongCollectivelyEncodedRydberg2021}. This scope is also being pursued with trapped Rydberg atoms, allowing enhanced coherence times \cite{Kuzmich2013, meiTrappedAlkaliMetalRydberg2022}.

This recent progress highlights the potential of the strong coupling enabled by Rydberg superatoms to be exploited in areas beyond nonlinear optics. As such, it is an area of research that deserves attention in the coming years.


\ack

J.K. acknowledges support from the Deutsche Forschungsgemeinschaft (DFG) within the research unit FOR 2247. C.B. acknowledges support by the Deutsche Forschungsgemeinschaft (DFG) under Germany’s Excellence Strategy – EXC-2111 – 390814868. N.S. and S.H. work has been supported by the European Union under ERC consolidator grant RYD-QNLO (grant N. 771417) and by the Carlsberg Foundation through the Semper Ardens project QCool. A.P-M. acknowledges support from CONACyT A1-S-29630 and and LN-321104. 

\bibliographystyle{iopart-num}
\bibliography{biblio}

\end{document}